# Generation expansion planning in the presence of wind power plants using a genetic algorithm model

Ali Sahragard [1], Hamid Falaghi [2], Mahdi Farhadi [3], Amir Mosavi [4,*] and Abouzar Estebsari [5]

[1] Khorasan Regional Electricity Company, Birjand, Iran; a.sahragard@krec.ir
[2] Department of Electric and Computer Engineering, University of Birjand, Iran; falaghi@birjand.ac.ir
[3] Department of Computer and Industries, Birjand University of Technology, Birjand, Iran; mahdifarhadi@birjandut.ac.ir
[4] Department of Informatics, J. Selye University, 94501 Komarno, Slovakia
[5] School of the Built Environment and Architecture, London South Bank University, London, United Kingdom; estebsaa@lsbu.ac.uk
* Correspondence: amir.mosavi@mailbox.tu-dresden.de

**Abstract:** One of the essential aspects of power system planning is generation expansion planning (GEP). The purpose of GEP is to enhance construction planning and reduce the costs of installing different types of power plants. This paper proposes a method based on Genetic Algorithm (GA) for GEP in the presence of wind power plants. Since it is desired to integrate the maximum possible wind power production in GEP, the constraints for incorporating different levels of wind energy in power generation are investigated comprehensively. This will allow obtaining the maximum reasonable amount of wind penetration in the network. Besides, due to the existence of different wind regimes, the penetration of strong and weak wind on GEP is assessed. The results show that the maximum utilization of wind power generation capacity could increase the exploitation of more robust wind regimes. Considering the growth of the wind farm industry and the cost reduction for building wind power plants, the sensitivity of GEP to the variations of this cost is investigated. The results further indicate that for a 10% reduction in the initial investment cost of wind power plants, the proposed model estimates that the overall cost will be minimized.

## 1. Introduction

Generation expansion planning (GEP) aims to find the most economically feasible solution to install a combination of multiple power generation in the long-term planning process of the power system. GEP determines the capacity, timing, and technology of new generation plants to provide the required energy for a 10- to 30-year time horizon. The increasing power consumption of industrial development, especially in developing countries, significantly highlights the need for GEP [1]. In one hand, considering the diverse and growing sources of renewable energy, wind power has valuable advantages, such as the capability to generate large quantities of cheap electricity, availability over a wide geographical area, and the possibility to create integrated wind-solar hybrid units. These factors have increased the importance of using this type of power plant. On the other hand, the use of wind power with intermittent behavior in production would significantly increase the complexities of the conventional GEP in which only thermal power plants are considered. GEP provides a mathematical model for an integer and constrained nonlinear optimization problem in which the objective function is satisfied if the constraints are met. To solve this optimization problem, two general mathematical and conceptual methods are used in [1]. The goal is to obtain a simple model that can be used in GEP studies. In other types of power plants, the output at this design level is considered a fixed number; however, in wind farms, it is necessary to obtain output as probabilities, and on the other hand, it should be as simple as possible so that it can be used in GEP.

In the wind farm model, it is assumed that all farms will be under a specific regime at a time. With this assumption, the simplified wind farm model is obtained.

There are currently new conceptual optimization methods, such as the particle swarm optimization (PSO) algorithm [2], the genetic algorithm (GA) [3], the honey bee algorithm [4], the taboo search (TS) [5], and ant colony (AC) [5], compared to mathematical methods such as linear programming (LP) [6], dynamic programming (DP) [7], and integer programming (IP) [8]. In addition to the variety of GEP optimization problem-solving techniques, the objective functions and the network-imposed constraints widely vary among different design cases. For example, objective functions can be profit maximization of a generation company in the restructured power market [9], maximizing reliability [10],[11], minimizing operating costs [12],[13], and minimizing environmental pollution [14],[15]. In addition, constraints such as network security constraints [16],[17],[18], [19] , [20] , [21], investment costs [22],[23],[24], reliability [25],[26],[27],[28],[29], and environmental pollution [30] could be part of the constraints that are required to be considered in the GEP problem. Although less attention has been paid to the GEP problem in the presence of wind power plants in the literature, extensive studies have been carried out on GEP of traditional power plants with various objective functions and constraints [31],[32]. From one side, the cost reduction of the initial investment in wind power plants over the past decade has led to a remarkable amount of renewable energy investment specifically allocated to this type of power plants [33],[34]. From the other side, the power generation capacity of these units is significantly lower than the nominal amount due to some new challenges posed by the presence of these intermittent generations in the power grid [35]. Planning for the expansion and operation of wind power plants for long-term intervals is a way to minimize these challenges. The use of the fast start-up power plants studied in [36] is another solution to reduce the vulnerability of the power system subject to the increasing penetration of wind power generation.

GEP studies for wind power plants require proper modeling of wind turbines and wind farms. Numerous models of wind power plants have been presented in various papers [37],[38]. In [39], the planning of combining generation units in the presence of wind units is investigated. In [40], for a short period of time, the costs of operating and investing of wind farms are investigated in addition to the traditional power generation units. In [41], the penetration of wind power generation of different designs for various short-term economic incentives is examined. Another study aimed at minimizing the amount of carbon produced in GEP [42]. Few studies of GEP in the last decade have focused on the costs of wind power plant investment in various projects. In this paper, the main objective is to investigate GEP in the presence of wind farms for a long-term target period. For this purpose, a model of the turbine and the wind farm is presented for long-term studies. The turbines are assumed to be HAWT. The model considers the turbine's Forced Outage Rate (FOR), which is used as input for GEP studies. The mathematical model of GEP problem is presented by defining the objective functions and constraints. And finally, a proposed GEP model is solved using the Genetic Algorithm optimization method. The impact of decreasing initial investment due to the growth of wind units' technology is demonstrated and discussed. Moreover, considering the importance of the maximum possible use of wind energy in the generation system, the maximum possible use of wind power in the GEP process subject to the constraints is investigated. It is worth noting that the impact of wind regimes on long-term planning studies using two different weak and strong wind regimes is illustrated in this paper.

Despite studies for wind power plants in GEP in the last decade, very limited number of studies have examined the cost of investment for various projects in system development planning. Concequently, the purpose of the present study is to investigate GEP with the presence of wind farms for a long time.

The difference between the present article and most of the extensive studies in this field is as follows:
- The objective function of this paper is to minimize the sum of expansion costs by considering the 4 constraints of maximum unit capacity to build, refueling constraints, storage margin and Loss of Load Probability (LOLP).

- In this paper, in addition to traditional power plants, the presence of wind power plants with a random generation nature is considered.
- Due to the growth of technology for building wind farms, the initial investment required to build these units has decreased. In the present paper, the effect of this price reduction on the influence of wind units on the generation system for a long period of planning is studied.
- Given the importance of the maximum possible use of wind energy in the production system, the maximum possible use of wind power in the GEP process is investigated, provided that the constraints are met.
- The impact of the wind regime on long-term planning studies using two different wind regimes has been shown due to the development of wind unit technology and the increasing reliability of these units as well as different wind regimes. The type of wind regime is selected for 2 sample cities in Iran. Type 1 wind regime as a weak wind regime and type 2 wind regime as a strong wind regime have been used to obtain wind turbine output.

In the following section, Section 2, turbine and wind farm modeling are discussed. Section 3 presets the GEP mathematical model, including the objective function and constraints as well as the Genetic Algorithm method for optimization and problem-solving. In Section 4, case studies based on real network data are shown by performing three experiments. These experiments include comparing the combined use of traditional power plants and wind farms with the ones with only traditional power plants, computing the maximum possible penetration of wind power plants, and computing the objective function sensitivity to the initial investment cost changes. Finally, in Section 5, the results of this study are summarized with some remarks.

## 2. Modeling

The model of system load adequacy assessment in the presence of wind farms has presented in [43]. As can be seen, in order to obtain the output of the wind farms, wind conditions must be applied as inputs to a farm which is containing a large number of wind turbines. GEP assumes that all loads and production units are assembled in one bus. Therefore, during planning, the location of units is not discussed and studies are performed at HL1 level. For reliability studies, a completely reliable transmission network is assumed. Each wind farm contains a large number of wind turbines. To obtain the wind farm model, one must first obtain the power output model for the wind turbines and then combine the model of these turbines to create the final model equivalent to the farm output.

### 2.1. Wind Turbine Model

The output characteristics of wind turbines are very different from those of the other turbines. Conventional power plants are capable of generating nominal output at all times of the year (with the exception of out-of-service times). However, wind turbines, in addition to the outage due to breakdowns, are sometimes unable to generate power due to wind speed dependency. As a result, wind turbine power output is a function of wind speed, and there is a nonlinear relationship between wind turbine output power and wind speed. The mathematical Equation for the power-velocity dependency is as Equation (1), as follows:

$$P_{out} = \begin{cases} 0 & ; \quad 0 \leq v \leq v_{cin} \\ (A + Bv + Cv^2) & ; \quad v_{cin} \leq v \leq v_r \\ P_r & ; \quad v_r \leq v \leq v_{co} \\ 0 & ; \quad v \geq v_{co} \end{cases} \quad (1)$$

In Equation (1), $v$ is the wind speed variable, $v_{cin}$ (cut-in wind speed) the minimum wind speed required to operate the turbine, $v_{co}$ (cut-out wind speed) is the maximum wind speed terminates turbine power generation, $v_r$ (rated wind speed) is the velocity of the nominal power of the turbine, and $P_r$ is the nominal power of the turbine. According to [43], in the Equation (1), the constants A, B, and C depend on the values of $v_{cin}$ and $v_r$ and are defined in Equations (2), (3), and (4) as follows.

$$A = \frac{1}{(v_{cin} - v_r)^2}\left[v_{cin}(v_{cin} + v_r) - 4(v_{cin} \times v_r)\left(\frac{v_{cin} + v_r}{2v_r}\right)^3\right] \quad (2)$$

$$B = \frac{1}{(v_{cin} - v_r)^2}\left[4(v_{cin} \times v_r)\left(\frac{v_{cin} + v_r}{2v_r}\right)^3 - (3v_{cin} + v_r)\right] \quad (3)$$

$$C = \frac{1}{(v_{cin} - v_r)^2}\left[2 - 4\left(\frac{v_{cin} + v_r}{2v_r}\right)^3\right] \quad (4)$$

*2.2. Wind Farm Model*

Although wind power generation has been considered adaptive to the environment, it has become an important issue due to the variable nature of this energy and its impact on system generation. Unlike traditional sources, the wind is not always available. Changes in wind energy output have created technical problems for the operation and generation expansion planning. In addition to being variable in the wind, changing the power output of wind farms, the unavailability of the turbine units can also lead to a change in power output. Therefore, the use of wind energy depends on the structure and weather conditions [44]. High utilization of wind farms creates fluctuations in the overall generation of the system that can vary depending on the amount of penetration in the system and the regional wind regime [44].

Wind farms can supply large amounts of energy, depending on the number of turbines installed. These farms can be connected to distribution and transmission networks. Different types of wind turbines with different capacities and output characteristics can be used in wind farms. To obtain the wind farm model, one must first obtain the power output model for the wind turbines and then combine this model to calculate the final model equivalent to the farm output. The wind sample used to obtain the wind turbine output was two-year wind data of Ardebil with an average velocity of 4 ($m/s$) [38]. In this section, the output power of a 2 MW unit is divided into six parts 0, 0.4, 0.8, 1.2, 1.6 and 2 that the turbine output model is obtained from Table 1 based on the data of the last two years, adapted from [45].

The number of scenarios considered for the output of a wind farm depends on the available data, the nature of the wind regime, the characteristics of the wind data, the computation time and the accuracy required [44]. Each wind farm contains a large number of wind turbines. Assuming the similarity of all these units, the power output of a farm for use in long-term studies of system planning can be obtained in two ways.

**Table 1.** Probability of turbine output power levels

| Probability | Output Power (MW) |
|---|---|
| $P(p_w < 0.2) = 0.4750$ | 0 |
| $P(0.2 \leq p_w < 0.6) = 0.3036$ | 0.4 |
| $P(0.6 \leq p_w < 1) = 0.0854$ | 0.8 |
| $P(1 \leq p_w < 1.4) = 0.0623$ | 1.2 |
| $P(1.4 \leq p_w < 1.8) = 0.0098$ | 1.6 |
| $P(1.8 \leq p_w) = 0.3036$ | 2 |

A farm contains a large number of turbines. Their output depends on the wind speed. All wind power plants in a wind farm are exposed to a wind regime with the same speed and characteristics, and the output model is calculated from Equation (5).

$$P = A \times X \quad (5)$$

In Equation (5), P is the equivalent output power vector of the combination of units (MW), A is the number of units, and X is the output power vector of each wind unit (MW), and the probability of each wind farm output mode is equal to the probability of the same state in the wind farm [44]. However, in our study, the Forced Outage Rate (FOR) of each wind turbine is considered in the wind farm modeling.

Forced Outage probability of each wind turbine is equivalent to FOR, and each turbine has a K-mode output model, as shown in Table 2, and the wind farm contains N turbine numbers.

Table 2. Output table of a turbine.

| Capacity (MW) | Probability |
|---|---|
| $p_1$ | $P_{WTG,1}$ |
| $p_K$ | $P_{WTG,K}$ |

Suppose $P_i$ is the probability of $i$ being the unit of power output and define as follows.

$$P_i = \binom{N}{i}(1 - FOR)^i FOR^{(N-i)} \tag{6}$$

Since each turbine has a number of $K$ output power levels, while the number of units available is $i$ units, the capacity available for each level of output of the turbines ($CAP_{avail}$) using Equation (7) is obtained, and the probability of this capacity is available to $P_{avail}$ can be calculated from Equation (8) as also described in [46].

$$CAP_{avail} = i \times P_{WTG,j}\, j = 1, \dots, K \tag{7}$$

$$P_{avail} = P_i \times P_{WTG,j}\, j = 1, \dots, K \tag{8}$$

The process of obtaining the wind farm output model is shown in Table 3.

Table 3. The process of obtaining an output model for a wind farm.

| Number of Units Available | Number of Units Exited | Available Probability ($P$) | Available Capacity ($CAP_{avail}$) | Possibility to Access $CAP_{avail}$ Capacity ($P_{avail}$) |
|---|---|---|---|---|
| 0 | N | $P_1 = N.1.FOR^{(N-0)}$ | $0 \times p_1$ | $P_0 \times P_{WTG,1}$ |
|   |   |   | $0 \times p_K$ | $P_0 \times P_{WTG,K}$ |
| i | N − i | $P_i = \binom{N}{i}(1 - FOR)^i FOR^{(N-i)}$ | $i \times p_1$ | $P_i \times P_{WTG,1}$ |
|   |   |   | $i \times p_K$ | $P_i \times P_{WTG,K}$ |
| N | 0 | $P_N = \binom{N}{N}.(1 - FOR)^N$ | $N \times p_1$ | $P_N \times P_{WTG,1}$ |
|   |   |   | $N \times p_K$ | $P_N \times P_{WTG,K}$ |

For a farm with N turbine unit, and K output mode for each turbine, the farm output model has (N + 1) × K different states. To simplify the scenarios, we have to classify the model based on the output power. Table 4 shows the outputs classified for a farm consisting of 30 turbine units presented in Table 2.

Table 4. Sixty-megawatt farm output model for FOR = 0.1.

| Probability | Output Power (MW) |
|---|---|
| $P(CAP_{avail} < 6) = 0.475$ | 0 |
| $P(6 \leq CAP_{avail} < 18) = 0.304265$ | 12 |
| $P(18 \leq CAP_{avail} < 30) = 0.089295$ | 24 |
| $P(30 \leq CAP_{avail} < 42) = 0.061224$ | 36 |
| $P(42 \leq CAP_{avail} < 54) = 0.028845$ | 48 |
| $P(54 \leq CAP_{avail}) = 0.041371$ | 60 |

## 3. Methodology

The mathematical model of the GEP problem consists of two parts: the objective function and the constraints as follows:

*3.1. Objective Functions*

GEP is divided into two-year periods, each of which is a planning stage. For each design stage, the objective function of the problem involves minimizing two different cost segments in Equation (9) as follows [47]:

$$min\, O.F = CapitalCosts + OperationalCosts \quad (9)$$

Reference [47] illustrates an adptation of the cost of different parts of the objective function.

*Capital Costs* include two components, *Investment Cost* and *Salvation Value*. In addition, *Operational Costs* is consisting of two components as *Fixed and Variable Operation & Maintenance Cost* and *Expected Energy Not Supplied* .Components of the objective function are described separately:

A. Investment Cost

*Investment Cost*, calculated per kW and has varying amounts for different types of units. This includes the cost of generating equipment and electrical equipment, the cost of building a fuel storage tank, the cost of connecting to the grid, and the cost of filters and equipment that are used for reducing environmental pollution.

In planning, this cost is assumed at the beginning of the stage in Equation (10) as follows[47]:

$$I(U_t) = (1+d)^{-t'} \cdot \sum_{i=1}^{N} [CI_i \cdot U_{t,i}] \quad (10)$$

The row matrix $U_t$ contains the capacity of the units added at the *t*-th stage of the planning and contains the planning-decision variables. If the number of units is equal to $N$ of a unit type, the vector $U_t$ at each step *t* contains N element in Equation (11) as follows:

$$U_t = (u_t^1, \dots, u_t^i, \dots, u_t^N) \quad (11)$$

where $u_t^i$ is the capacity of units of type *i* to be constructed in the *t*-th phase of the planning.

Furthermore, in Equation (11), $d$ is the interest rate, $CI_i$ the initial investment cost required for units of type *i* ($/MW) and $u_t^i$ the capacity of new units added of type *i*, in the *t*-th step is used. The parameter $t'$ is used to transfer the invested costs at the beginning of each planning step to the base year and is defined in Equation (12) as follows:

$$t' = t_0 + s \times (t-1) \quad (12)$$

*s* is the number of years considered for each step, which, as mentioned, is often 2 years for planning.

B. Salvation Value

After the plant is built, over time and considering the unit Exhaustion Rate, at the end of its life, the unit's equipment and facilities still have a value equivalent to a percentage of the purchase cost. The amount that can be calculated from Equation (13) is the *unit residual value*, the *capital value of the residual* or the *salvage value of the unit*:

$$S(U_t) = (1+d)^{-T'} \cdot \sum_{i=1}^{N} [\delta_{t,i} CI_i \cdot U_{t,i}] \quad (13)$$

In Equation (13), $\delta_{t,i}$ is the *cost-return factor* for unit type *i*. Since the *residual value of the units* is considered at the end of each planning step, parameter $T'$ is used to transfer it to the base year, which is defined in Equation (14) as follows:

$$T' = t_0 + s \times T \quad (10)$$

C. Fixed and Variable Operation and Maintenance Cost

Fixed operation and maintenance cost per MW is computed over a month or a year and includes costs related to overhaul, maintenance, tax and employees' payroll costs. In addition to that, variable operation and maintenance cost, covers the cost of energy supplied by the units is obtained during each design step per KWh. Since energy production is proportional to fuel consumption per unit, this cost is equivalent to the fuel cost which is required. Therefore, the total operating cost is defined in Equation (15) as follows:

$$M(X_t) = \sum_{y=0}^{s-1}\left[(1+d)^{-(t'+0.5+y)} \times \sum_{i=1}^{N}[X_{t,i} \times FC_i + MC_i \times EES_{t,i}]\right] \quad (15)$$

It should be noted that the vector $X_t$ is a cumulative vector of $U_t$ as Equation (16):

$$X_t = X_{t-1} + U_t \quad (t = 1, \dots, T) \quad (16)$$

In Equation (15), $X_{t,i}$ is the capacity of existing units of type $i$ in the $t$-th period, $FC_i$ is constant operating cost of type $i$ ($/MW), $MC_i$ is the variable cost of operating unit type $i$ at the $t$-th stage ($/MWh) and $EES_{t,i}$ is the amount of energy that unit type $i$ provides at the $t$-th period. Since the operating cost is routinely spent during each phase (not at the beginning or end of the phase), according to [47], using $y$ parameter, the costs are transferred to the middle of the year for each year and then transferred to the base year for use in the objective function. Moreover, to determine the variable costs, the energy provided by each unit at each planning stage $EES_{t,i}$ approximately calculates the area under the *Load Duration Curve* (LDC).

Since generation expansion planning is performed for long periods, accurate prediction of the LDC curve is not possible and only the maximum load at each design stage is available, so the LDC curve approximation has been used based on an adaptation from [47].

The base load for each step is considered as a percentage of the peak load of that period. To calculate the amount of energy supplied by each unit at each stage ($EES_{t,i}$), at each planning stage, the units are arranged in descending order of variable operating cost. Using the LDC linear curve, the energy supplied by each unit, which is equivalent to the area below the curve, is calculated from Equation (17).

$$\text{if } \text{Time} = f(\text{Load}) \text{ then } EES_{t,i} = \int_{L1}^{L2} f(\text{Load}) \times d\text{load} \quad (11)$$

If the horizontal axis is the time axis and the vertical axis is the load, in Equation (17) f(Load) in the LDC curve is the characteristic of the load function of each stage in term of time. dload the differential value of the load used to calculate the surface area provided by each power unit.

L1 is the level of generation capacity before adding $i$-th unit capacity and L2 is the level of generation capacity after adding $i$-th unit capacity. However, in the generation expansion planning instead of using a nonlinear LDC, a two-piece equivalent linear model is used and the above integral is easily calculated[47].

The amount of energy supplied by each unit at each step ($EES_{t,i}$) is equivalent to the area below the LDC curve.

D. The *Expected Energy Not Supplied* (EENS) Cost

In terms of the importance of the reliability of the system being planned, its cost is considered in the operating costs of the objective function. One of the parameters for reliability measurement of the system is the Expected Energy Not Supplied which is equivalent to the Outage Cost during the planning period. The cost of Energy Not Supplied per MWh of energy during each planning stage is obtained from Equation (18):

$$O(X_t) = \sum_{y=0}^{s-1} \left[ (1+d)^{-(t'+0.5+y)} . EENS_t . CEENS \right] \tag{18}$$

In Equation (18), $EENS_t$ of energy not supplied in the *t*-th stage of planning (MWh) and $CEENS$ is the value of each MWh of energy ($). As can be seen in [47], the load is lost for $t_k$ if the system output capacity is $Q_k$ and is greater than the Reserve Margin. In this situation, $p_k$ is the probability that $Q_k$ is out of capacity.

In this case, $EENS$ is defined as Equation (19)

$$EENS = \sum_{Q_k > ReserveMargin} S_k \times p_k \tag{19}$$

In Equation (19), $S_k$ is the amount of energy that is lost in the system if $Q_k$ capacity outage occurs. Assuming the linearity of the LDC curve, the colored area ($S_k$) can be easily obtained [47].

Thus, from the all of the above studies, the ultimate objective functions of the GEP problem are to minimize the sum of initial investment costs, operating, system outage, or not supplied energy value, and the residual value of the units in Equation (20) as follows:

$$minO.F = \sum_{t=1}^{T} \left( I(U_t) + M(X_t) + O(X_t) - S(U_t) \right) \tag{20}$$

*3.2. Constraints*

In general, the constraints governing the GEP problem can be divided as the below [47]:

A. Practical Constraints

The types of constraints associated with the process of adding units to the system are called executive constraints. These constraints include:

The maximum number of units that can be manufactured

There are restrictions on the number of units to be built at each planning stage for executive reasons. This constraint is considered in Equation (21) as follows:

$$0 \leq U_t \leq U_{t.max} \tag{21}$$

$U_{t.max}$ is the vector of the maximum capacity of new possible units for the programming stage *t*.

*A.1. Fueling Constraint*

Various types of units with different fuels such as furnace oil, natural gas, coal and nuclear fuel can be considered in the GEP. Given this constraint, system decision-makers can choose to combine generation to reduce the risk of dependence on a particular type of fuel. This constraint can be considered as Equation (22):

$$M_{min}^i \leq \frac{X_{t,i}}{\sum_{k=1}^{N} X_{t,k}} \leq M_{max}^i \tag{22}$$

In Equation (22), $M_{min}^i$ and $M_{max}^i$ are the minimum and maximum ratios of the type *i* unit used in the *t*-th stage of planning, respectively.

*A.2. Pollution*

The increasing importance of environmental protection requires power plants to comply with relevant laws and standards in the design process to reduce the amount of unit pollution in the form of system constraints. In some cases, even pollution and its costs are considered as part of the objective function. In this article, the pollution-related constraint is ignored.

B. Technical Constraints

These constraints must be met by analyzing the system at each planning stage to ensure that the system reliability is at an acceptable level. In general, system reliability evaluation is divided into two subsets of System Adequacy and System Security. What is considered in long-term GEP is the System Adequacy section. System Adequacy is about having sufficient facilities to power customers at all times, which checks for planned and unplanned outages of system elements. System Adequacy can generally be divided into Probabilistic and Deterministic methods [48].

*B.1. Reserve Margin constraint*

In the deterministic method, Reserve Margin is a definite criterion used to evaluate system reliability by specifying the Generation Margin. The Generation Margin is the percentage of surplus capacity installed at the annual peak load obtained from Equation (23) [48]:

$$ReserveMargin = \frac{InstalledCapacity - PeakLoad}{PeakLoad} \times 100\% \quad (23)$$

If $\sum_{k=1}^{N} X_{t,k}$ is the total installed capacity of the system in stage *t*, including existing and new units, any acceptable design shall meet the following condition as Equation (24):

$$(1 + R_{min}) \times D_t \leq \sum_{k=1}^{N} X_{t,k} \leq (1 + R_{max}) \times D_t \quad (24)$$

where $R_{min}$ and $R_{max}$ are the minimum and maximum system reservations, respectively. $D_t$ is also the maximum predicted load for the programming stage *t*.

*B.2. Loss of Load Probability (LOLP) Constraint*

If the electrical system is more complex and larger in size, the Reserve Marginal one will not be sufficient to assess reliability. The deterministic method, which uses only Reserve Margin calculations, results in overinvestment or insufficient reliability. However, the *Reserve Margin* in GEP can be as high as 15% to 40%, because GEP is usually done for more than a decade, with forecasting times associated with error and therefore, high percentages are viewed as Reserve Margin.
The major disadvantage of the deterministic method is that it does not respond to the random and probabilistic nature of system behavior, customer demand, and system component error, and the system risk is not determined by this method. The probabilistic method for evaluating system adequacy, which has been in use since 1930s, provides a comprehensive overview of the probability set of probabilistic events and examines system reliability indices [48].
Among the various probability parameters of reliability, the Loss of Load Probability index (LOLP) as a constraint has to be considered in the GEP problem.
If the capacitance $Q_k$ is lost, the load is lost for $t_k$ and in this case, $p_k$ is the probability of outage capacitance $Q_k$, LOLP can be obtained as follows [47]:

(25)

$$LOLP = \frac{\sum_{Q_k > ReserveMargin} p_k \times t_k}{T}$$

In the GEP problem, the units selected together with the new units must meet the LOLP criteria in Equation (26) as follows:

$$LOLP(X_t) \leq \varepsilon \quad (26)$$

where ε is the maximum allowed value of LOLP.

*3.3. GA Optimization*

In this paper, the GA algorithm was used to solve the GEP optimization problem [46-55]. The general trend of the GA algorithm for solving the GEP problem is shown in Figure 1. The algorithm uses integer coding to form genes. Each gene can have an integer from zero to the maximum number of constructible units in each programming step. For the situation involving T the planning stage and selection from the N possible unit types in each period, the chromosomes would be as Equation (27):

$$U = (U_1, U_2, \ldots, U_l, \ldots, U_T) \qquad (27)$$

In which the capacity of the selected units $U_l$ of each of the $N$ types of power plant units in the *l*-th design stage is

$$U_l = (u_l^1, u_l^2, \ldots, u_l^N) \qquad (28)$$

The algorithm is implemented in such a way that all the constraints of the system are satisfied while selecting the gene values for each programming stage and thus, the chromosomes production process of each generation goes out of pure random selection.

**4. Data**

*A. Traditional Power Generation Units*

The technical and economic information of the existing generation system, as well as the candidate unit specifications, are consistent with the ones provided in Tables III and IV in [49], respectively. This system has 15 production units installed with different capacities. The total installed capacity is 5100 MW.

*B. Wind Farm*

The nominal capacity of each wind farm is 60 MW, including 30 turbines with the power of each turbine equal to 2 MW. The six-state model output power of 60 MW and 2 MW wind units used for two different wind regimes are shown in Table 5 and Table 6, respectively. Wind regime 1 is based on Ardebil wind data [38]. The $v_{cin}$, $v_r$ and $v_{co}$ speeds of the turbine used are 4, 15 and 25 m/s, respectively [45]. Wind regime 2 was also obtained for a windy region (Figure 1). The fixed part of the cost of operating and maintaining wind farms is 11500 \$per MW in a year. The variable cost of operating and maintaining wind farms is \$ 0.0025 per kWh. The initial investment cost per kW of wind farms is also \$ 1485 [11].

Table 5. Sixty-megawatt wind farm output power model for two wind regimes.

| Probability for Wind Regime 1 | Probability for Wind Regime 2 | Output Power (MW) |
| --- | --- | --- |
| 0.2942 | 0.475 | 0 |
| 0.174601 | 0.304265 | 12 |
| 0.165499 | 0.089295 | 24 |
| 0.184267 | 0.061224 | 36 |
| 0.096124 | 0.028845 | 48 |
| 0.084879 | 0.041371 | 60 |

Table 6. Two-megawatt wind farm output power model for two wind regimes.

| Probability for Wind Regime 1 | Probability for Wind Regime 2 | Output Power (MW) |
|---|---|---|
| 0.2942 | 0.475 | 0 |
| 0.1734 | 0.3036 | 0.4 |
| 0.1543 | 0.0854 | 0.8 |
| 0.1694 | 0.0623 | 1.2 |
| 0.07714 | 0.0098 | 1.6 |
| 0.1314 | 0.0639 | 2 |

The six-state output model of a wind farm, including 30 turbines with the power of each turbine equal to 2 MW was obtained in accordance with the described method. In both regimes, for each 2 MW unit, FOR is set at 0.1. The output characteristic of a wind turbine in wind regime 2 (strong wind regime) is shown in Figure 1. Furthermore, figure 2 compares these two regimes.

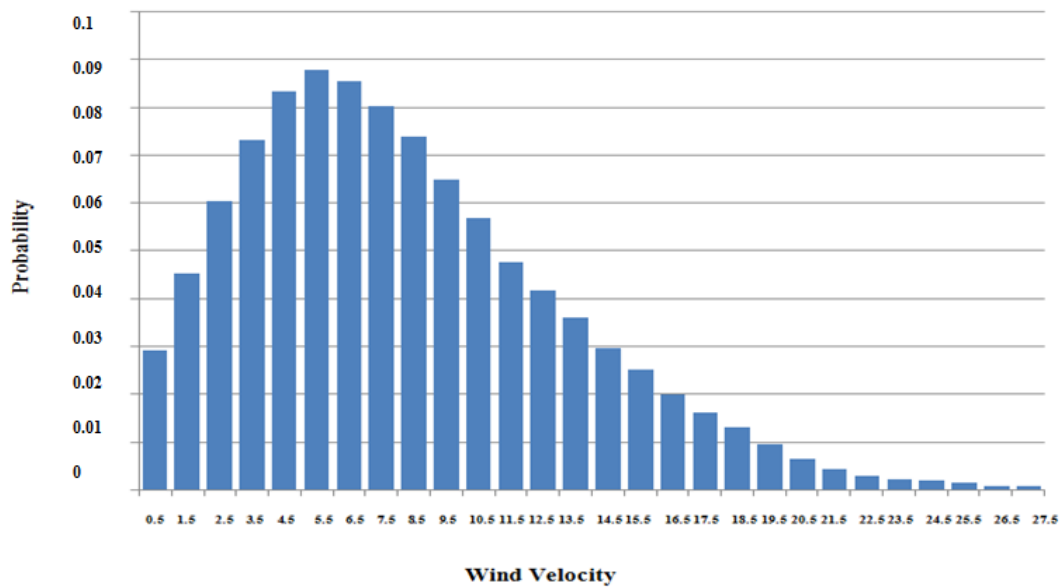

**Figure 1.** Output characteristic of a wind turbine in wind regime 2 (strong wind regime).

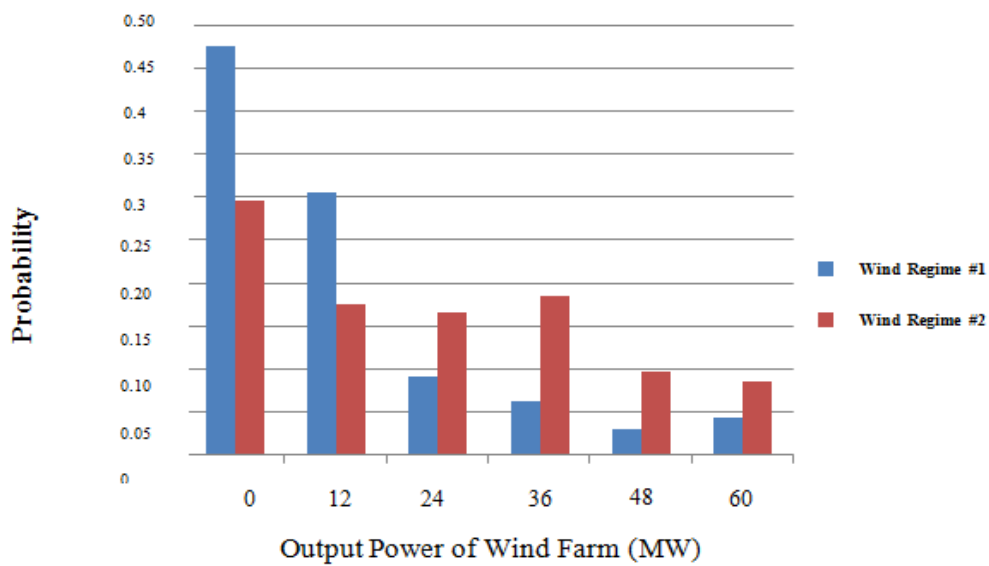

**Figure 2.** Histogram of 60 MW wind farm output power for two wind regimes.

*C. Forecast Load*

The forecasted load for the next 14 years of the system is assumed as shown in Table 7 which is consistent with the assumption in [49]. A two-piece linear LDC curve is used to calculate the energy during the planning, and the baseload is considered to be 50% of the peak load of each planning stage.

Table 7. Forecast peak load.

| Planning Stage | 0 | 1 | 2 | 3 | 4 | 5 | 6 | 7 |
|---|---|---|---|---|---|---|---|---|
| Forecasted Load (MW) | 5000 | 7000 | 9000 | 11000 | 13000 | 15000 | 17000 | 19000 |

*D. Objective Function and Constraint Parameters*

The interest rate (d) in the study was 8.5%. The EENS cost of 0.05 $/ was is also included in the studies. The maximum LOLP value is 0.01. The minimum and maximum reservations are 15% and 40%, respectively.

The remaining capital coefficient values for OIL, LNG, COAL, PWR, PHWR and WIND units are 0.1, 0.1, 0.15, 0.2, 0.2 and 0.1, respectively. The minimum and maximum fuel mix ratio of the various units is shown in Table 8.

Table 8. A fuel mix ratio of units used in planning.

| Fuel Mix Ratio | Max (%) | Min (%) |
|---|---|---|
| OIL | 0 | 30 |
| LNG | 0 | 60 |
| COAL | 20 | 60 |
| PWR | 30 | 60 |
| PHWR | 30 | 60 |

The interval between the study and the beginning of planning ($t_0$) was two years. The GEP problem was implemented for seven planning stages in which no wind farm case should have an optimal combination of the generation system with five types of power plants; and in the case of a wind farm, the number of units must be chosen from among the six types of power plants so that the objective function is (at minimum) optimal.

*E. Specifications of GA*

When only studies are performed in the presence of traditional units (five types of units), for the seven programming stages, each chromosome contains 35 genes. In Section 4.2, which also includes the number of wind units in the chromosomes forming a single selectable type, each chromosome will have 42 genes. Integer coding is used to form genes. That is, each gene can have an integer value from zero to the maximum number of constructible units at each programming stage.

To reproduce the generation in each iteration, the crossover operator was used for 60% of the population. The Roulette wheel is used to select the type of crossover and the probability of each type of one-point, two-point and substring crossovers is 0.7, 0.15 and 0.15, respectively. The mutation operator was used for one generation percentage, equivalent to three chromosomes per generation. To maintain the best chromosomes of each generation, the three chromosomes of each generation that have the best value for the objective function are transferred unchanged to the next generation. The initial population consists of 300 chromosomes, and GA was repeated for 150 generations. Of course, to get the best possible answers, every GA was run multiple times and the best answer was reported as the optimal answer.

## 5. Case Studies

In this paper, two types of wind regime, including type 1 wind regime as the weak wind regime for Ardebil city in Iran, and type 2 wind regimes as a strong wind regime for a windy sample city, were used to obtain wind turbine output [38]. The output power of a 2 MW turbine unit was divided into 6 parts, 0, 0.4, 0.8, 1.2, 1.6 and 2 MW and for FOR = 0.1, using the output model calculation procedure according to Equations (2)–(4), the output models classified for two farms consists of 30 turbine units calculated and is shown in Table 9.

**Table 9.** Sixty-megawatt wind farm output power model for two types of wind regimes.

| Output Power (MW) | Probability for Wind Regime 2 | Probability for Wind Regime 1 |
|---|---|---|
| 0 | 0.2942 | 0.475 |
| 12 | 0.174601 | 0.304265 |
| 24 | 0.165499 | 0.089295 |
| 36 | 0.184267 | 0.061224 |
| 48 | 0.096124 | 0.028845 |
| 60 | 0.084879 | 0.041371 |

The required technical and economic data of the traditional power plants and wind farms studied in this article have been extracted entirely from references [49] and [11], respectively. In Section 4, the technical and economic information of the system under study was introduced. The studies in this article are presented in the form of three experiments as follows:

*5.1. GEP in the presence of wind farm*

In the first experiment, the planning was performed separately for two separate schemes. In plan 1, a 14-year planning interval, including seven 2-year planning stages, was implemented for the traditional power plant units. In plan 2, wind units were also considered as a selectable type of unit in the planning process. Thus, for plan 1, only studies were performed in the presence of five traditional unit types, for 7-stage planning, each chromosome contains 35 genes, whereas for plan 2, which also includes the wind units as a selectable unit type in chromosome formation, each chromosome will have 42 genes.

For wind farms, the minimum number of constructions in each period is one unit, and the initial investment cost of wind farms is $ 1485 per kWh. The results of the two selected optimal plans and the costs of each plan are presented in Tables 10 and 11.

**Table 10.** Selected optimal plans with the least cost function, in the presence of wind farm (plan 1) and in the presence of a wind farm (plan 2).

| Power Plant | OIL | | LNG | | COAL | | PWR | | PHWR | | WIND | |
|---|---|---|---|---|---|---|---|---|---|---|---|---|
| Plan | 1 | 2 | 1 | 2 | 1 | 2 | 1 | 2 | 1 | 2 | 1 | 2 |
| Stage 1 | 3 | 0 | 4 | 2 | 0 | 1 | 0 | 1 | 0 | 0 | 0 | 1 |
| Stage 2 | 3 | 0 | 2 | 2 | 1 | 2 | 0 | 0 | 1 | 0 | 0 | 2 |
| Stage 3 | 0 | 0 | 3 | 4 | 1 | 1 | 0 | 0 | 1 | 1 | 0 | 1 |
| Stage 4 | 3 | 1 | 1 | 1 | 1 | 2 | 1 | 1 | 0 | 0 | 0 | 3 |
| Stage 5 | 4 | 2 | 0 | 2 | 1 | 1 | 0 | 0 | 1 | 1 | 0 | 3 |
| Stage | 2 | 1 | 0 | 0 | 1 | 1 | 1 | 0 | 0 | 2 | 0 | 3 |
| 6 | | | | | | | | | | | | |
| Stage 7 | 0 | 3 | 1 | 1 | 1 | 1 | 1 | 1 | 0 | 0 | 0 | 2 |

**Table 11.** Selected optimal design with the least cost function.

| Optimal Plan | 1 | 2 |
|---|---|---|
| Total Cost (M$) | 17136,679 | 17245,513 |
| Investment Cost of Initial Plan (M$) | 11461,397 | 11468,975 |
| Operational Cost of Plan (M$) | 3772,192 | 3723,465 |

The results show that due to the high cost of initial investment of wind farms compared to other units, the total cost of investment increases, although the use of wind farms in planning leads to reduce the cost of operation.

*5.2. Investigating the impact of wind penetration on GEP*

The scenario which was considered in the second experiment was aimed at investigating the impact of adding wind to the system in a step-by-step manner. A certain number of fixed wind farm steps were added to the system at each stage. For example, to achieve 4% wind penetration in the generation system, two wind units must be constantly added to the production system at each planning stage, to the end of the 7-stage planning of 14 wind farm installations that have a capacity equivalent to 4% of the total peak load available in the generation system. To achieve 9, 13, 18, and 22 percent wind penetration, 4, 6, 8, and 10 wind farms must be constructed in each 2-year design period, respectively. Since the number of wind units added in this scenario is initially fixed and constant, the number of chromosome genes is 35. Only in the process of problem-solving, the effect of the added wind units at each stage should be considered in the constraints of the problem. The process of adding stairwells continues until the LOLP constraint was violated. This scenario was implemented for both types of wind regime, and the results were obtained.

Figure 3 illustrates the cost function changes as the wind farm penetrates the system. Although using wind farm as shown in Figure 4 reduces the Operational Cost of the system due to its low operating cost compared to other types of power plants, since the cost of building these units is high, adding these units to the production system increases the overall cost of the project.

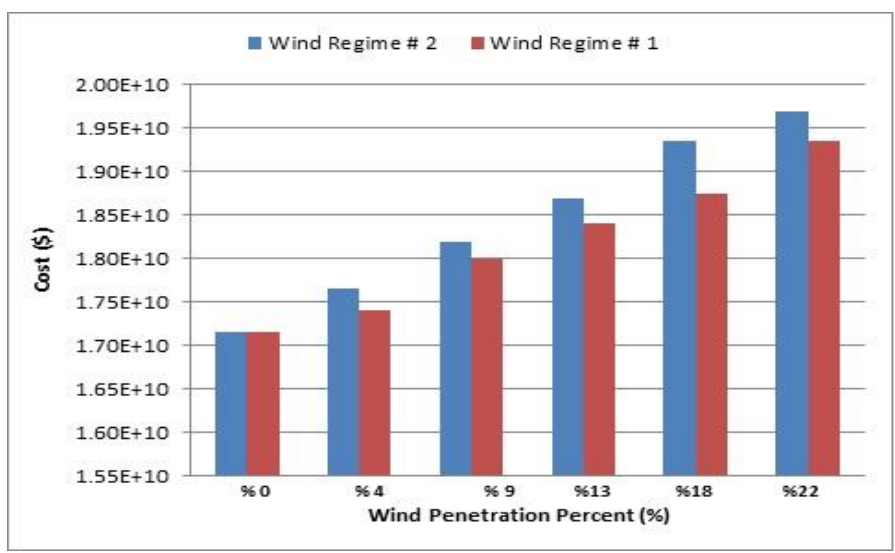

**Figure 3.** Increasing the cost of wind penetration plans in the system.

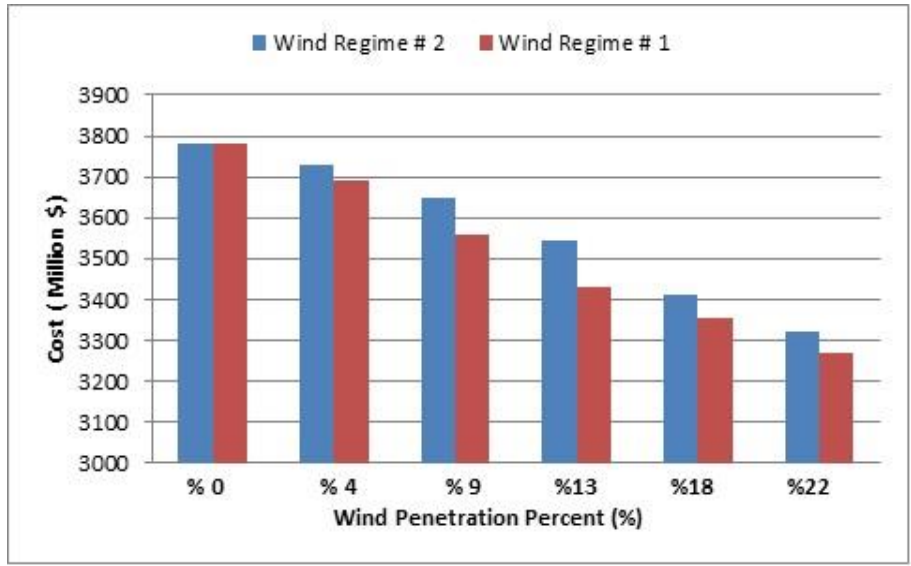

**Figure 4.** Reducing the operational cost of wind penetration plans in the system.

The effective amount of wind farm output power is the expected value of farm output power. For wind regimes 1 and 2 this is 11.8652 and 22.4075 MW, respectively. Due to the small cost of variable operating and maintenance of wind farms, these units were used in the process of calculating the operating cost to support the load. However, due to the low effective power output, the impact of increasing the cost of building these units was far greater than the cost of operating these units and increases the overall cost of the project.

The effective amount of wind farm output with wind regime 1 is less than regime 2. For this reason, the amount of reduction in operating costs for regime 2 is higher than that of regime 1.

As mentioned, the addition of wind farms can continue as long as the LOLP constraint is not violated, even if the increased cost incurred by the system is used. For the studied system, the LOLP constraint was violated in exchange for 22% penetration of wind farms with type 1 wind regime. Therefore, it is possible to install wind farms in the system if the wind regime type is one, with 8 farms at each planning stage. Figure 5 shows the different values of LOLP at different planning stages for different wind permeation values with the Type 1 regime.

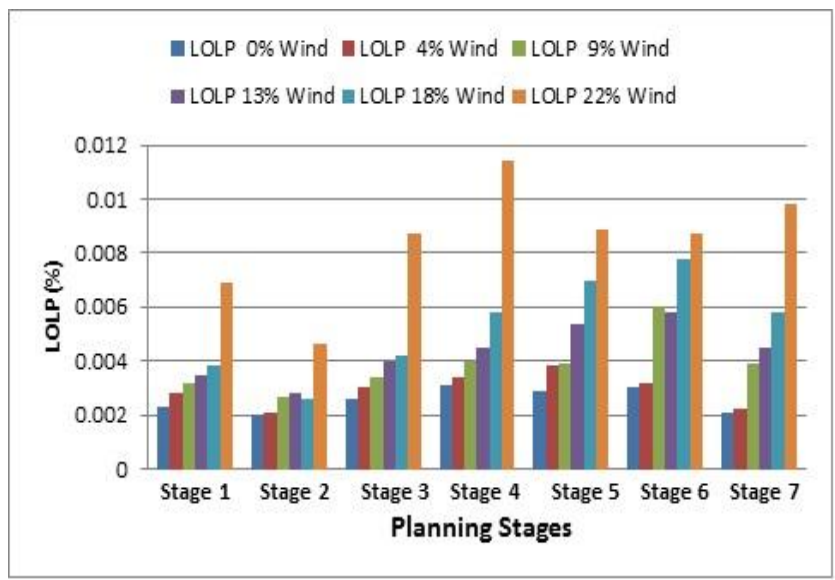

**Figure 5.** Changes in loss of load probability (LOLP) plans with different type 1 wind regimes.

As shown in Figure 6, for the penetration of regime-1 wind farms on the generation system, the LOLP constraint approaches the limit value of 0.01, and for the 22% penetration of this constraint in the 4-th planning stage, it was violated. However, if the LOLP constraint of Figure 5 is examined for wind regime 2, it is found that for the 26.5% wind penetration in the system, this constraint is violated due to the more appropriate wind regime 2 because of the higher reliability of the ratio to the wind regime is 1. Hence, the maximum possible installation of a wind farm with 2 wind regimes at each planning stage is 10 farms.

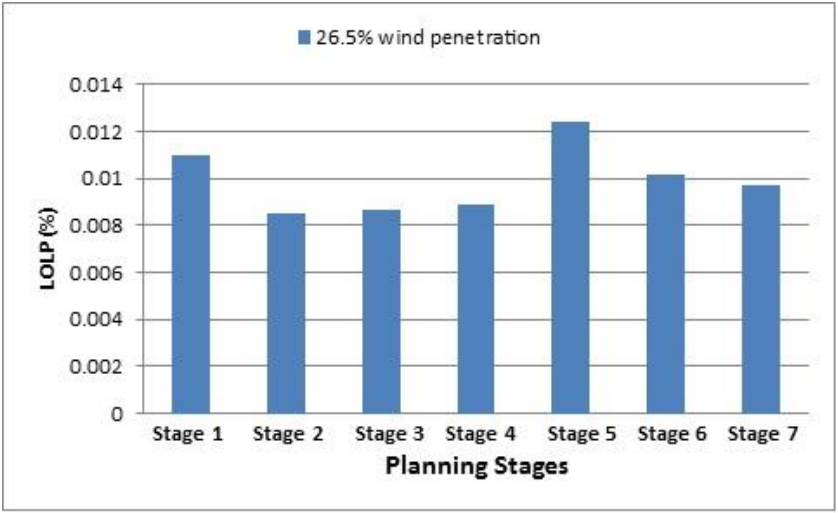

**Figure 6.** LOLP constraint violation for 26.5% wind penetration in the system.

*5.3. Investigating the Sensitivity of the Problem of Initial Investment in Wind Farms*

In the third experiment, wind farms are considered as one of the types of power plant units to be selected. The minimum number of wind farm options for both types of wind regime is 1 unit and the maximum for type 1 wind regime is 8 units and for wind 2 regimes is 10 units. This experiment was conducted to determine the sensitivity of the objective function to the initial investment cost of the wind units using two turbine output power models derived from two different wind regimes. For this purpose, for each of the two initial investment amounts less than $ 1485 (1402 and 1320, respectively), and for the higher initial investment amounts of 1575 and 1650, different design costs are compared. As shown in Figure 7, by reducing the initial investment cost of the wind units, the cost of the required designs is reduced and by increasing the investment cost of these units, the total cost of the optimal design increases.

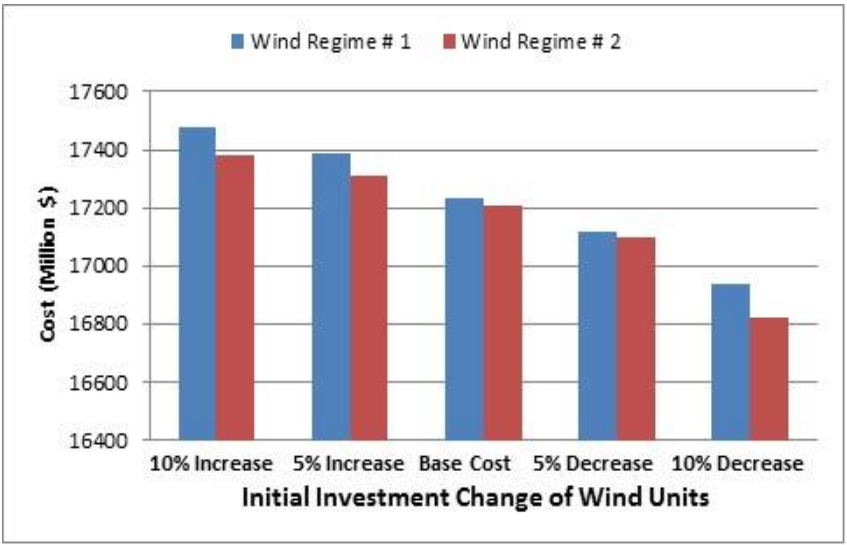

**Figure 7.** The impact of reduced wind unit investment on the objective function.

When the initial investment cost of wind units ($/KW) is 1650, the number of wind units in the optimally selected plan will yield the minimum designated number one. By reducing the cost of initial investment in optimizing the layout, more wind farms are being selected. Interestingly, in order to reduce the cost of investing wind farms to ($/kW) 1320, the final design selected has a lower cost than one that is not used in any wind farm system. In other words, if the cost of initial investment required for wind farms is reduced to less than ($/kW) 1320, plans are selected that, despite the use of wind farms, cost less than the total cost of not using wind farms.

In the system under study, a plan with 8% wind penetration was chosen as the optimal design for reduction of the initial investment required for wind farms to ($/kW) 1320 for type 1 regime and for farms with wind regime 2, the selected plan has 11% wind penetration. Figures 8 and 9 illustrate the amount of variation in initial investment cost and operating cost of the selected designs for different amounts of initial unit investment cost reduction. Figure 10 shows the change in the number of units selected in different plans for regime 2.

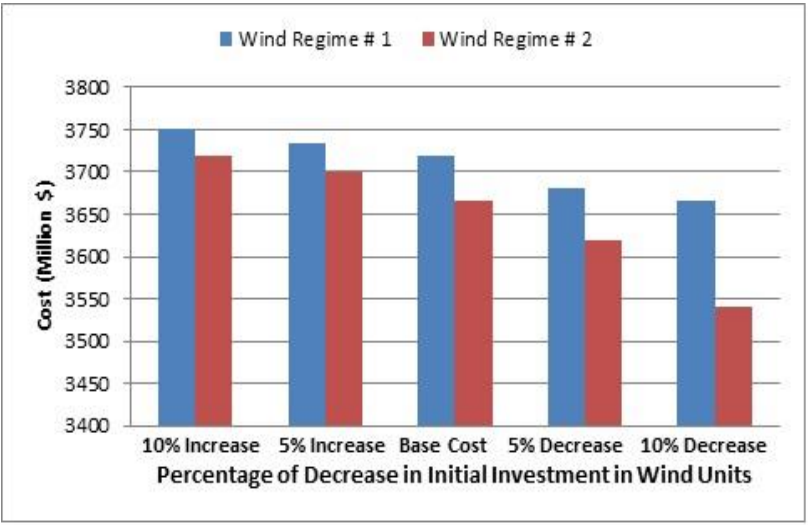

**Figure 8.** Cost reduction of system operation by increasing wind farm penetration.

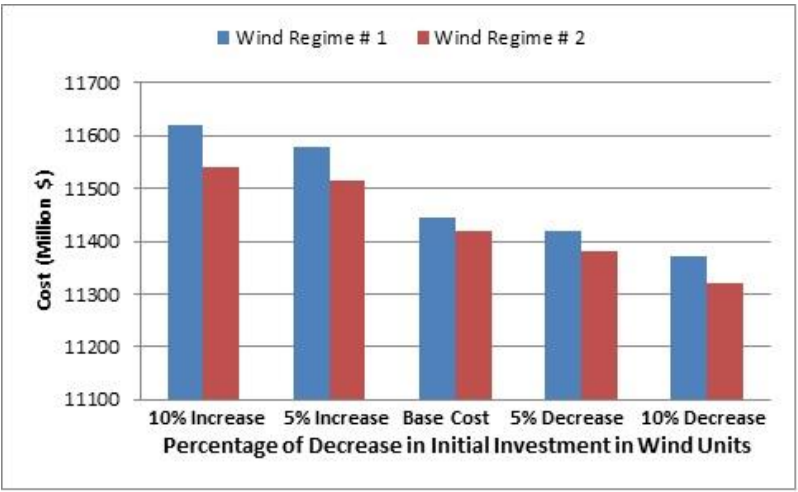

**Figure 9.** The impact of reducing the cost of investing wind farms in initial system investment.

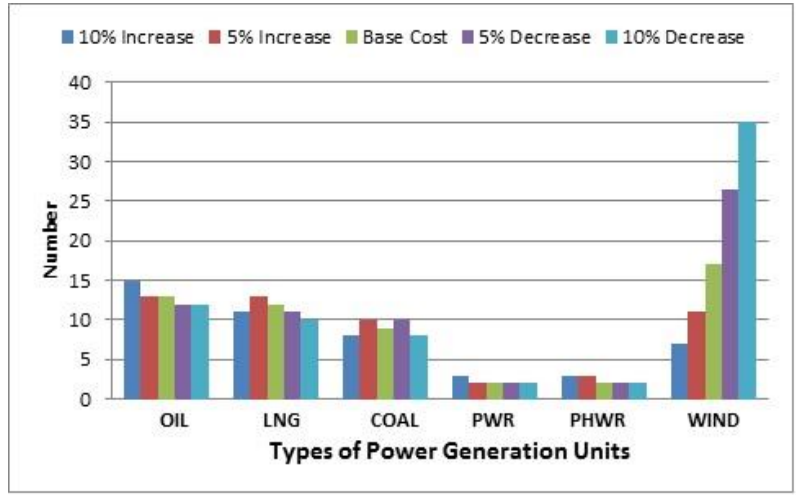

**Figure 10.** Change in the number of optimal design units by changing the cost of investing in regime 2 wind farms.

Due to the number of OIL units selected in different plans, it can be said that with the increase in the possibility of using wind units, the system is moving towards less use of OIL units. Due to the cost of investment required for OIL units, it is determined by reducing the initial investment cost of wind farms and making the price competitive with OIL units, given the lower operating costs of the wind units, plans use less OIL units.

## 6. Conclusions

In this paper, Genetic Algorithm was used to solve the problem of GEP. A six-state model was used to obtain the wind farm output power model. The method of calculating the six-state wind farm output model with the Forced Outage Rate (FOR) of wind farm units for use in long-term GEP calculations is described. Next in the first experiment of GEP using genetic optimization algorithm (GA) an optimized plan was obtained. In the second experiment, depending on the importance of knowing the maximum utilization of wind farms in plans for system planning, the maximum possible penetration was calculated by increasing the number of wind units in steps. Due to the existence of different wind regimes in different regions, two models of power output were developed for strong and weak wind regimes. The outcomes of the models were studied and compared. It was observed that by increasing the capacity utilization of wind farms, the cost of the plans increases. This was less for areas with a strong wind regime than for areas with a weak wind regime.

In addition, by installing wind farms in areas with strong wind regimes, the maximum amount of capacity which is available is increased, subject to all constrains. Due to the growth of technology associated with the construction of wind farms and the reduction in construction cost per kW of wind, the sensitivity of the objective function to the change in construction cost was tested in the third experiment. It was found that for a 10% reduction in cost, the construction of these units can be found in a combination of generating units that, while using wind farms, cost less than the total plan cost.

Finally, the studies presented in this paper show that by reducing the cost of initial investment due to the improvement of wind turbine and wind farm technology, the competitive potential of wind power plants is significantly increased compared to other power plants. This justifies the increase in the number and capacity of wind power plants. It is worth mentioning that the problem of generation expansion planning was investigated with some assumptions that by revising these assumptions we can re-examine the effect of these factors. The following can be suggested to complement the research:

- A model with a higher number of scenarios for the wind farm can be used to increase the accuracy of the calculations.
- In this study, the predicted two-piece linear model was used, while a more accurate model can be used to make the results more realistic.
- Uncertainties in both forecasted load and costs can also be included in the calculations and its effect can be examined.
- Given the problem with the HL1 level, the transmission system is assumed to be quite reliable, which can be considered the transmission network.
- Models can be used to simultaneously consider wind, solar, and other types of renewable energy sources and address the issue.
- In the issue of generation expansion planning, which was examined in this study, only the type of unit, the number and time of units being added to the final design are specified.
- Running the problem by considering the network can determine the location of the units and the impact of the location on the reliability of the system. In this case, different regimes of wind can be applied to different regions, making the results closer to reality.
- This study is conducted as a single bus, therefore the network is not considered. At the next level of system planning so-called Transmission Expansion Planning (TEP), the network should be studied, hence the role of increasing reactive power and low power factor due to the expansion of wind power can be investigated.

**Acronyms**

| | | | |
|---|---|---|---|
| GEP | Generation Expansion Planning | $P_{avail}$ | probability of the $CAP_{avail}$ |
| FOR | Forced Outage Rate | P | Available Probability |
| LOLP | Loss of Load Probability | Capacity ($P_{avail}$) | Possibility to Access $CAP_{avail}$ |
| $v$ | wind speed variable | K | output mode for each turbine |
| $v_{cin}$ | (cut-in wind speed) minimum wind speed required to operate the turbine | $O.F$ | Objective Function |
| $v_{co}$ | (cut-out wind speed) maximum wind speed terminates turbine power generation | $I$ | Investment Cost |
| $v_r$ | (rated wind speed) is the velocity of nominal power of the turbine | $U_t$ | capacity of the units added at the $t$-th stage of the planning |
| $P_r$ | nominal power of the turbine | $u_t^i$ | capacity of units of type $i$ to be constructed in the $t$-th phase of the planning |
| P | equivalent output power vector of the combination of units (MW) | $d$ | interest rate |
| A | the number of units | $CI_i$ | initial investment cost required for units of type $i$ ($/MW) |
| X | output power vector of each wind unit (MW) | $s$ | number of years considered for each step, which, is often 2 years for planning |
| $P_i$ | probability of $i$ being the unit of power output | S | Salvation Value |
| K | number of output power levels | $\delta_{t,i}$ | cost-return factor for unit type i |
| $i$ | number of available units | $T'$ | parameter is used to transfer to the base year |
| $X_t$ | Cumulative vector of $U_t$ | $EENS_t$ | energy not supplied in the $t$-th stage of planning (MWh) |
| $X_{t,i}$ | capacity of existing units of type $i$ in the $t$-th period | $CEENS$ | value of each MWh of energy ($) |
| $FC_i$ | constant operating cost of type $i$ ($/MW) | $t_k$ | Time which the system output capacity , |
| $MC_i$ | variable cost of operating unit type *i* at the *t*-th stage ($/MWh) | | $Q_k$ is greater than the Reserve Margin and load is lost |
| $EES_{t,i}$ | amount of energy that unit type *i* provides at the *t*-th period | $p_k$ | probability that $Q_k$ is out of capacity |
| $y$ | operating cost routinely spent during each phase (not at the beginning or end of the phase) | $O$ | the Outage Cost during the planning period |
| | | $S_k$ | amount of energy that is lost in the system if $Q_k$ capacity outage occurs |
| LDC | Load Duration Curve | $U_{t.max}$ | vector of the maximum capacity of new possible units for the programming stage *t* |
| F (Load) | characteristic of the load function of each stage in term of time in the LDC curve | $M^i_{min}$ | minimum ratios of the type *i* unit used in the *t*-th stage of planning, respectively |
| L1 | level of generation capacity before adding *i*-th unit capacity | $M^i_{Max}$ | maximum ratios of the type *i* unit used in the *t*-th stage of planning |
| L2 | level of generation capacity after adding *i*-th unit capacity | dload | the differential value of the load used to calculate the surface area provided by each power unit |
| dload | the differential value of the load used to calculate the surface area provided by each power unit | $R_{min}$ | minimum system reservations |
| $R_{max}$ | maximum system reservations | $p_k$ | probability of $Q_k$ |
| $D_t$ | maximum predicted load for the programming stage *t* | ε | maximum allowed value of LOLP |
| $Q_k$ | outage capacitance , capacitance   is lost, | | |

**Author Contributions:** Conceptualization, H.F.; Data curation, A.S.; Formal analysis, A.S. and H.F.; Investigation, H.F.; Methodology, A.S. and H.F.; Project administration, H.F.; Resources, A.S., H.F. and M.F.; Software, A.S.; Supervision, H.F.; Validation, A.S., A.M. and H.F.; Visualization, H.F.; Writing—original draft, M.F. and A.E.; Writing—review and editing, M. F., A.M. and A.E.; funding acquisition, A.M.

**Funding:** This research received no external funding.

**Conflicts of Interest:** The authors declare no conflict of interest.